\documentclass[aps,twocolumn,10pt,prl,floatfix,nofootinbib, superscriptaddress]{revtex4-2}

\usepackage{amsmath} 
\usepackage{amssymb}
\usepackage{bm} 
\usepackage[retainorgcmds]{IEEEtrantools}
\usepackage{graphicx}
\usepackage{mathrsfs}
\usepackage{color} 
\usepackage{amsfonts}
\usepackage{tikz}
\usepackage[colorlinks=true,linkcolor=RoyalBlue,urlcolor=RoyalBlue,citecolor=RoyalBlue]{hyperref}
\usepackage{lipsum}
\usepackage{soul}
\usepackage{physics} 
\usepackage{float}
\usepackage{dsfont} 
\usepackage{mathrsfs}
\usepackage{BOONDOX-ds}
\usepackage{marvosym}
\usepackage{nicefrac}
\usepackage{bbold}
\usepackage{cancel}
\usepackage[version=4]{mhchem}
\usepackage[english]{babel}

\definecolor{mint}{RGB}{164, 215, 210} 
\definecolor{rot}{RGB}{210, 5, 55}
\definecolor{lightrot}{RGB}{235, 130, 155} 
\definecolor{antri}{RGB}{45, 55, 60}
\definecolor{RoyalBlue}{cmyk}{1, 0.50, 0, 0}

\newcommand{\nb}{\bar{n}}
\newcommand{\bin}{b_\text{in}}
\newcommand{\dbin}{b_\text{in}^\dagger}
\newcommand{\bout}{b_\text{out}}
\newcommand{\dbout}{b_\text{out}^\dagger}

\begin{document}

\title{Bridging Quantum and Semi-Classical Thermodynamics in Cavity QED}
\author{Marcelo Janovitch} 
\email{m.janovitch@unibas.ch} 
\affiliation{Department of Physics and Swiss Nanoscience Institute,
	\\ University of Basel, Klingelbergstrasse 82, 4056 Basel,
Switzerland }
\author{Sander Stammbach}
\affiliation{Naturwissenschaftlich–Technische Fakultät, Universität Siegen, 
    \\ Walter-Flex-Straße 3, 57068 Siegen, Germany}
\author{Matteo Brunelli}
\affiliation{JEIP, UAR 3573 CNRS, Coll\`ege de France, PSL Research University,
11 Place Marcelin Berthelot, 75321 Paris Cedex 05, France}
\author{Patrick P. Potts} 
\affiliation{Department of Physics and Swiss Nanoscience Institute,
	\\ University of Basel, Klingelbergstrasse 82, 4056 Basel,
Switzerland }
\date{\today}

\begin{abstract}
In cavity quantum electrodynamics (QED), photons leaving the cavity can be irreversibly lost or reused as a power source. This dichotomy is reflected in two different thermodynamic bookkeepings of the light field, both corresponding to valid thermodynamic frameworks. In this work, we formulate a rigorous semi-classical limit of cavity QED and show that the resulting thermodynamic description may qualitatively differ from that of the fully quantised model. We find that violations of the thermodynamic uncertainty relations are recovered in the semi-classical limit only by one of the two thermodynamic frameworks: the one which treats part of the photon flux as a power source. We illustrate our findings in a three-level system coupled  to a driven cavity.
\end{abstract}

\maketitle

\begin{figure}[t]
    \centering    \includegraphics[width=0.5\textwidth]{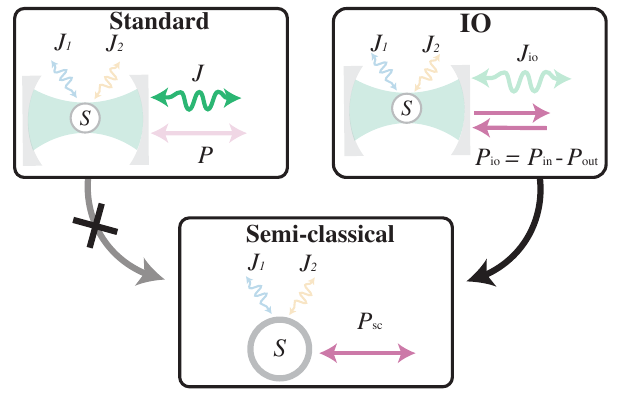}
    \caption{Sketch of the general results. Wiggly (straight) arrows indicate incoherent (coherent) fields. The standard thermodynamic framework treats the coherent part of the output as waste, while the input-output framework (IO) as useful energy. The standard framework leads to predictions which are incompatible with the semi-classical model, while the IO-framework leads to compatible ones. 
    }
    \label{fig:results_sketch}
\end{figure}

\textit{Introduction.---}Quantum devices typically operate out of equilibrium: they continuously produce entropy and sustain stationary currents.  
These currents fluctuate due to both classical noise and genuinely quantum effects. The \emph{thermodynamic uncertainty} associated with a current $I$ is captured by the dimensionless quantity (with $\hbar=k_\text{B}=1$)
\begin{align}
    \mathcal{Q}(I,\sigma)=\frac{\langle\!\langle I^2\rangle\!\rangle}{\langle I\rangle^2}\,\sigma,
    \label{eq:Q}
\end{align}
where $\langle I\rangle$ and $\langle\!\langle I^2\rangle\!\rangle$ are the mean and scaled variance of the current, respectively, $\sigma$ is the entropy-production rate, and all quantities are taken in the long-time limit. Bounds on $\mathcal{Q}$, known as \emph{thermodynamic uncertainty relations} (TURs), have become central tools in non-equilibrium statistical physics~\cite{Barato2015, Gingrich2016, Pietzonka2016, Pietzonka2017, Horowitz2020, Falasco2020}. 
The canonical bound $\mathcal{Q}\ge 2$ holds for classical Markovian dynamics, whereas quantum coherence may suppress fluctuations and thus violate it. Such violations are, however, highly model dependent~\cite{Saryal2019, Janovitch2023, Agarwalla2018, Liu2019, Cangemi2020, Menczel2021, Ehrlich2021, Prech2023, Ptaszyski2023,  Almanza2025}. 

Cavity quantum-electrodynamics (QED) platforms~\cite{Goban2014, Tiecke2014, DeSantis2017, Luan2020, Takahashi2020, Wang2019, Najer2019, Janovitch2024} offer exquisite control over light--matter interactions and are natural candidates for implementing quantum thermal machines. 
However, the thermodynamic analysis of cavity QED systems is subtle, owing to the coherent coupling  to a quantised cavity field, which is itself driven and dissipative. Directly applying known expressions for the thermodynamic bookkeeping~\cite{Alicki_1979, spohn1978} to a cavity QED system leads to what we refer to as the \textit{standard} thermodynamic approach~\cite{Boukobza2006_bipartite, Vinjanampathy_2016,Potts_2021, potts2024quantumthermodynamics}. In this approach, photons leaving the cavity are uniquely associated to a heat current and therefore contribute to the total entropy production. An alternative approach for coherently driven systems, based on \textit{input--output theory} (IO), has recently been formulated~\cite{schrauwen2025,2024_prasad}. In this framework, the coherent part of the outgoing photons does not contribute to the entropy production. The rationale is that the coherent output field of the cavity can be used as a power source for another quantum system.
While both approaches provide consistent thermodynamic frameworks—each equipped with a first law of thermodynamics and a non-negative entropy production rate—they differ in how entropy production is quantified, resulting in different values of the thermodynamic uncertainty, Eq.~\eqref{eq:Q}. This is in contrast to a semi-classical description, in which the cavity field reduces to an external time-dependent parameter that does \textit{not} produce entropy. How the thermodynamics obtained from a semi-classical description relates to the different thermodynamic approaches to cavity QED is an open question.

Here, we answer this question by rigorously formulating a semi-classical limit of cavity QED systems and benchmarking the two thermodynamic approaches against this limit. 
Starting with a fully quantised model, we show that the standard thermodynamic approach does not recover the thermodynamic description of the semi-classical model, as the light keeps producing entropy in this limit. Instead, the input-output framework recovers the thermodynamic description of the semi-classical model, see Fig.~\ref{fig:results_sketch}. 

Our work clarifies the physical origin of the power output in the semi-classical descriptions of quantum thermal machines based on cavity QED: the power comes from the coherent part of the emitted field. 
Although this conclusion may seem  obvious from a quantum optics perspective, we show that it has major consequences for TURs, which we illustrate in the three-level maser~\cite{Geusic1967, kosloff_1984, Mitchison2019}, i.e., a thermally driven three-level atom coupled to a cavity.
In a semi-classical description, this system is known to violate the TUR due to the presence of quantum coherence~\cite{Kalaee_2021}. Upon treating the field quantum mechanically,  TUR violations only persist if the input-output approach to thermodynamics is employed. In the standard approach, the dissipation associated to the light overestimates the entropy production, preventing any TUR violations.

\textit{Setup.—}We consider a quantum system embedded in a driven–dissipative cavity. 
The Hamiltonian consists of the system, the driven cavity, and their interaction:
\begin{align}
H(t) = H' + H_0(t) + V,
\end{align}
where $H'$ is the bare system Hamiltonian and
\begin{align}
H_0(t) = \Omega a^\dagger a 
+ i\mathcal{E}\!\left(a^\dagger e^{i\omega_\text{d} t} - a e^{-i\omega_\text{d} t}\right),
\end{align}
describes a cavity with frequency $\Omega$ laser-driven with amplitude $\mathcal{E}$ and frequency $\omega_\text{d}$.
For concreteness we focus on an interaction of the form
\begin{align}
V = g(aO^\dagger + a^\dagger O),
\label{eq:flipflop-interaction}
\end{align}
where $O$ is a system operator; generalisations to multi-photon interactions are discussed in the Supplementary Material \cite{supmat}.

The composite cavity-system dynamics is governed by the master equation
\begin{align}
\dv{\rho}{t} = -i[H(t),\rho] + \mathcal{D}[a]\rho + \mathcal{D}'\rho,
\label{eq:lme}
\end{align}
where $\mathcal{D}$ describes a bath at temperature $T$ coupled to the cavity via 
\begin{align}
\mathcal{D}[a] 
= \kappa(\nb+1)D[a] + \kappa\nb D[a^\dagger],
\end{align}
with $D[L]\circ = L\circ L^\dagger - \tfrac12\{L^\dagger L,\circ\}$ and $\nb= [e^{\omega_\text{d}/T}-1]^{-1}$ the Bose–Einstein occupation of the cavity bath. For Markovian master equations, thermodynamic consistency requires using the appropriate frequency $\omega_\text{d}$ in the occupation $\nb$; note, however, that $[e^{\omega_\text{d}/T}-1]^{-1}\simeq [e^{\Omega/T}-1]^{-1}$ in order for Eq.~\eqref{eq:lme} to be justified~\cite{Wacker2022, Potts_2021}. Dissipative processes
acting on the system  are collected in $\mathcal{D}'$,
\begin{align}
\mathcal{D}' = \sum_{j=1}^r \Gamma_j D[L_j],
\end{align}
where the jump operators $L_j$ and rates $\Gamma_j$ are model-dependent.

Our analysis assumes a coherently driven cavity-QED system described by quantum Markovian dynamics. The results therefore rely on the standard assumptions underlying Lindblad master equations and quantum white-noise input fields: weak system–bath coupling, a flat bath spectrum, and the rotating-wave approximation~\cite{gardiner_1985, Breuer&Petruccione2002}. The intra-cavity system itself may be arbitrary and can include multi-photon interactions with the cavity mode \cite{supmat}. 


\textit{Semi-classical limit.—}We now identify the parameter regime in which the cavity field behaves as a classical coherent drive.  This is the case when the average cavity field dominates over its fluctuations; we then write
\begin{align}
a = \alpha(t) + \tilde{a},
\end{align}
where $\alpha(t)$ is the amplitude of the driven cavity in the long-time limit, and in the absence of coupling to the system (i.e., for $g=0$),
\begin{align}
\alpha(t) = -2\frac{\mathcal{E}}{\kappa} \chi e^{-i\omega_\text{d}t},
\qquad
\chi = \frac{1}{1+2i\Delta/\kappa},
\end{align}
with detuning $\Delta = \Omega - \omega_\text{d}$. 
Since $\alpha(t)$ corresponds to $\langle a\rangle$ in the limit $t\to\infty$, $g\to0$, the displaced mode generally satisfies $\langle\tilde{a}\rangle \neq 0$.

We define the \textit{semi-classical limit} as
\begin{equation}
|\alpha| \to \infty,
\qquad
\frac{g}{\kappa} \to 0,
\qquad
\frac{g}{\kappa}|\alpha| = \text{const.}.
\label{eq:sc-lim}
\end{equation}
In this regime the cavity hosts a large coherent field, whereas noise from thermal fluctuations and system back-action is negligible.  
Thus, $\tilde{a}$ represents only weak quantum fluctuations around the dominant classical amplitude. For multi-photon interactions, the coherent amplitude in the last expression 
of Eq.~\eqref{eq:sc-lim} must be rescaled accordingly~\cite{supmat}.
Upon taking this limit and tracing out the cavity mode, we obtain the \textit{semi-classical model}~\cite{supmat}:
\begin{align}
\dv{\rho'}{t}
&= -i[H_\text{sc}(t),\rho'] + \mathcal{D}'\rho', \label{eq:semi-classical-lme} \\ 
H_\text{sc}(t) &= H' + g\!\left[\alpha(t)O^\dagger + \alpha^*(t)O\right], \nonumber
\end{align}
where $\rho'$ is the reduced density matrix of the intra-cavity system.

Formally, connecting Eq.~\eqref{eq:lme} to Eq.~\eqref{eq:semi-classical-lme} follows from a controlled perturbative expansion in the small parameter $g/\kappa$, combined with enforcing the scaling in Eq.~\eqref{eq:sc-lim}. In this regime, the residual coupling between the system and the fluctuation operator $\tilde{a}$ contributes only at subleading order, while the coherent cavity amplitude remains finite in the effective drive. Dropping the terms containing $\tilde{a}$ in the reduced dynamics of the intra-cavity system is thus justified in the semi-classical limit (details in \cite{supmat}).
Physically,  Eq.~\eqref{eq:semi-classical-lme} reduces the cavity field to a classical amplitude: it no longer mediates back-action on the system and acts purely as an externally controlled, time-dependent drive. In contrast, the intra-cavity system remains quantum.

In the semi-classical model, the effective drive is directly associated with the injected \textit{power},
\begin{align}
    \langle P_\text{sc} \rangle = \langle \partial_t H_\text{sc} \rangle= -i \omega_\text{d} g  \,\langle\alpha(t) O^\dagger - \alpha^*(t)O\rangle,
\end{align}
while the dissipative currents flowing into the system via $\mathcal{D}'$ contribute to the entropy production rate,
\begin{align}
    \sigma_\text{sc} = -\sum_{j} \frac{\langle J_j \rangle}{T_j}.\label{eq:sigma-sc}
\end{align}
Explicit expressions for the system's heat current $\langle J_j \rangle$ are model-dependent and will be provided in the example below; in the general case they can be established using a thermodynamic Hamiltonian~\cite{supmat, schrauwen2025, Potts_2021}. The semi-classical model obeys the first and second laws of thermodynamics, i.e., $\langle P_\text{sc}\rangle +\sum_j\langle J_j\rangle = 0$ and $\sigma_\text{sc}\geq 0$.

Two important remarks follow. First, both the composite model Eq.~(\ref{eq:lme}) and the semi-classical model Eq.~(\ref{eq:semi-classical-lme}), give the same expression for the \textit{system} heat currents. Second, in the semi-classical model there is no contribution to the heat current coming from the cavity field, i.e., $\langle J_\text{sc} \rangle = 0$. 

\textit{Incompatibility of the standard framework.—}According to the standard thermodynamic framework applied to the composite system of Eq.~\eqref{eq:lme}, the average power is given by~\cite{Potts_2021},
\begin{align}
    \langle P \rangle= \langle\partial_t H(t) \rangle = -\omega_\text{d} \mathcal{E}\, \langle a e^{-i\omega_\text{d}t} + a^\dagger e^{i\omega_\text{d}t}\rangle.
\end{align}
The heat current from the cavity to its bath is given by,
\begin{align}
   \langle J\rangle &= \omega_\text{d} \tr\qty( a^\dagger a \mathcal{D} \rho) = \omega_\text{d} \kappa (\nb - \langle a^\dagger a \rangle). 
\end{align}
Together with the system's heat current $\langle J_j \rangle$, these two quantities satisfy the first law of thermodynamics $\langle P \rangle + \langle J \rangle + \sum_j \langle J_j \rangle =0$.
The entropy production is given by,
\begin{align}
       \sigma = -\frac{\langle J\rangle}{T} - \sum_j\frac{\langle J_j\rangle}{T_j},
\end{align}
which satisfies the second law: $\sigma\geq 0$.

We show in~\cite{supmat} that in the semi-classical, and long-time limit,
\begin{align}
        \langle P \rangle &= \langle P_\text{sc} \rangle -\langle J \rangle, \label{eq:P_to_sc}\\
    \langle J \rangle &= -\kappa \omega_\text{d}|\alpha|^2 - 4 \omega_\text{d} |\alpha| g \Im \langle O ~ e^{-i\omega_\text{d}t} \rangle |\chi|\label{eq:J_to_sc}.
\end{align}
\begin{align}
    \sigma = \sigma_\text{sc} - \frac{\langle J \rangle }{T},\label{eq:sigma_standard}
\end{align}
reflecting the failure of the standard approach to recover the thermodynamics of the semi-classical model.

Note that Eqs.~(\ref{eq:Q},~\ref{eq:J_to_sc}) imply that $\mathcal{Q}(J_j, \sigma) \not \to \mathcal{Q}(J_j, \sigma_\text{sc}) $ in the semi-classical limit. Importantly, this means that in the semi-classical limit $\sigma$ always diverges, due to the extensive contribution $\propto |\alpha|^2$ in Eq.~\eqref{eq:J_to_sc}.
Thus, the condition $\mathcal{Q}(J_j, \sigma)<2$, which would violate the classical TUR, cannot be reached, failing to witness non-classical effects in the fluctuations of $J_j$.

From a quantum-optical perspective, the extensive mismatch between  Eqs.~(\ref{eq:sigma-sc}) and (\ref{eq:sigma_standard}) in the semi-classical limit has a clear explanation. The coherent component of the cavity field 
can be absorbed into the Hamiltonian by a unitary displacement transformation, leaving the entropy unchanged. Equivalently, coherent and incoherent scattering play distinct thermodynamic roles: coherent outgoing radiation is work-like and may in principle be converted reversibly into useful work, for example using interferometric operations, whereas fluctuations constitute the entropy-producing part of the output field. This distinction is standard in quantum optics and is incorporated thermodynamically in the input--output formulation of Refs.~\cite{PhysRevLett.124.130601, schrauwen2025}. In the following, we use this framework to determine which thermodynamic bookkeeping of the fully quantised cavity-QED model is compatible with the semi-classical limit.

\begin{figure*}
    \includegraphics[width=\textwidth]{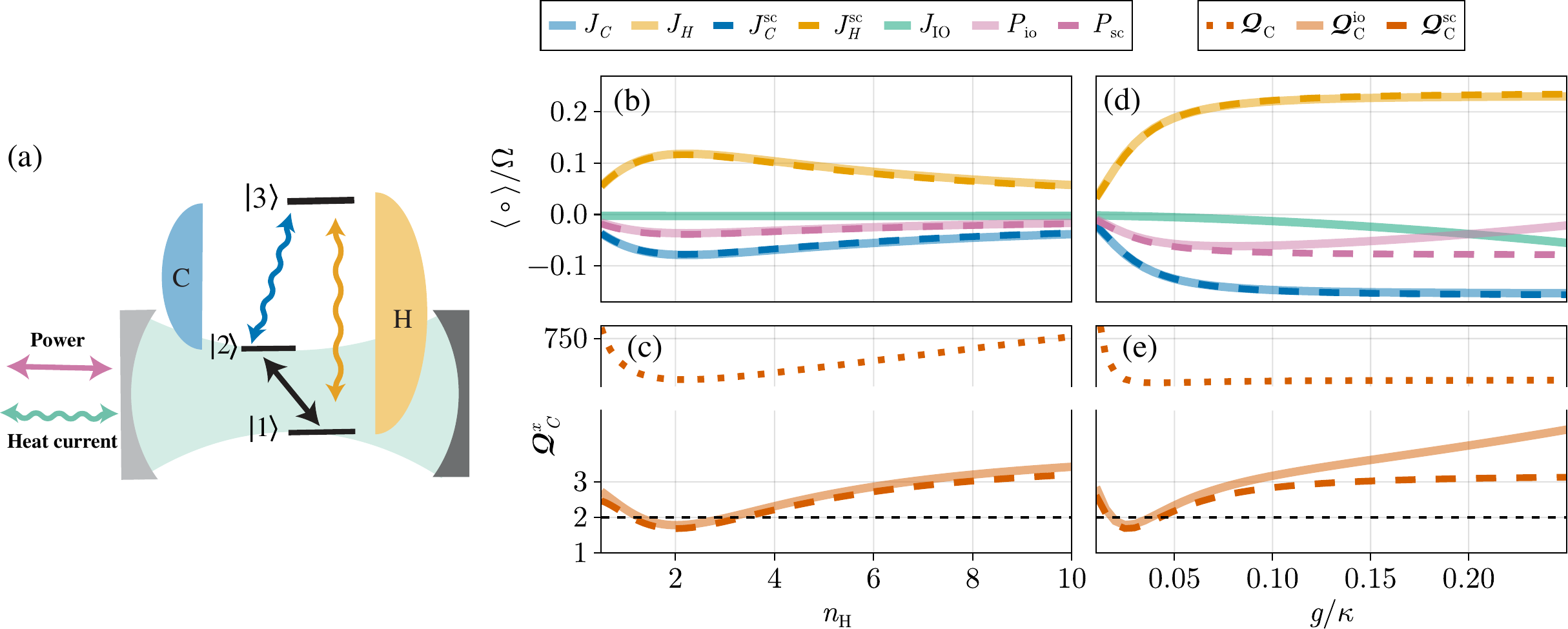}
    \caption{
    (a) Sketch of the cavity-embedded three-level maser coupled to hot (H) and cold (C) baths. 
    In (b,c) the hot-bath occupation $n_{\mathrm{H}}$ is varied (horizontal axis) at fixed $g/\kappa$. (b) Average heat currents, $\langle J_{H/C}^x \rangle$ for the three-level systems and for the cavity $\langle J^\text{io} \rangle$ and power $\langle P \rangle$. Averages for the standard framework result in large values $\propto \mathcal{E}^2/\kappa^2$ (not shown). According to the IO-framework, the cavity-embedded maser operates as a heat engine, in agreement with its semi-classical version. (c) Thermodynamic uncertainties of the cold heat current, $\mathcal{Q}_C^x,~x\in\{\varnothing,~\text{io},~\text{sc}\}$ for the dotted, transparent-solid, and solid orange curves, respectively. Semi-classical TUR violations are observed, showing consistency of the IO-framework, and inconsistency of the standard framework.
    In (d,e) $n_{\mathrm{H}}$ is fixed and the coupling $g/\kappa$ is varied. We observe the breakdown of the semi-classical limit for (d) average heat currents, power, and (e) thermodynamic uncertainties.
    In (b--e), fixed parameters are 
    $\mathcal{E}/\kappa = 1.5$, 
    $g/\kappa = 0.025$, 
    $\gamma_{\mathrm{H}}/\kappa = 0.1$, 
    $\gamma_{\mathrm{C}}/\kappa = 2.0$, 
    $\Omega/\kappa = \omega_2/\kappa = \omega_{\mathrm{d}}/\kappa = 3.5\times 10^3$, 
    $\omega_3 = 3\omega_2$, 
    $T/\kappa = 2\times 10^3$, 
    $T_{\mathrm{C}} = T$, and 
    $T_{\mathrm{H}}/\kappa \simeq 7.4 \times 10^4$. The mild value of  $\mathcal{E}/\kappa$ together with the low cavity temperature allows a cut-off for the cavity mode at $30$ levels,  leading to occupation probabilities $\propto 10^{-6}$ at $n=30$ \cite{Janovitch_QuTLM_numerics_2026}.}
    \label{fig:panel}
\end{figure*}

\textit{Compatibility of the input--output framework.—}
Compared to the standard approach, the IO framework redistributes the coherent contribution of the photons between power and cavity heat current. Specifically, it attributes to power the coherent contributions from the external field, and to heat its fluctuations; this is encoded in the definitions~\cite{PhysRevLett.124.130601, schrauwen2025},
\begin{align}
     \langle P_{\rm io} \rangle &=- \omega_{\rm d}\left(|\langle\bout(t)\rangle|^2 -|\langle\bin(t)\rangle|^2\right) \label{eq:Pio}\\
        & = P_\text{in} - P_\text{out}, \nonumber\\
        \langle J_{\rm io} \rangle &= -\omega_{\rm d}\left(\langle\!\langle\dbout(t)\bout(t)\rangle\!\rangle -\langle\!\langle\dbin(t)\bin(t)\rangle\!\rangle\right) \label{eq:Jio},
\end{align}
where $b_\text{in (out)}$ are the input (output) fields from input-output theory~\cite{gardiner_1985} and $\langle\!\langle AB\rangle\!\rangle = \langle AB\rangle-\langle A\rangle\langle B\rangle$. Heat currents  $\langle J_j\rangle$ are the same as above.
Since  the equality $\langle J \rangle + \langle P \rangle = \langle J_\text{io} \rangle + \langle P_\text{io} \rangle$ 
holds, the first law also follows~\cite{schrauwen2025}. Furthermore,
 \begin{align}
    \sigma_\text{io} = -\frac{\langle J_\text{io}\rangle}{T} - \frac{\langle J_\text{C}\rangle}{T_\text{C}} - \frac{\langle J_\text{H}\rangle}{T_\text{H}}.
 \end{align}
Importantly, $0\leq \sigma_\text{io}\leq \sigma$: a tighter second law~\cite{schrauwen2025}.

In the semi-classical and long-time limit, we find~\cite{supmat}
\begin{equation}
    \langle P_\text{io}\rangle = \langle P_\text{sc}\rangle\label{eq:Pio-scaling-maser},\hspace{1.5cm}
    \langle J_\text{io}\rangle = 0,
\end{equation}
which implies the main results,
\begin{align}
    \sigma_\text{io} &= \sigma_\text{sc}, \label{eq:Sigma-scaling-maser}\\
    \mathcal{Q}(J_j, \sigma_\text{io}) &= \mathcal{Q}(J_j, \sigma_\text{sc}).
\end{align}
Below, we illustrate the importance of utilising the thermodynamic uncertainty compatible with the semi-classical limit in assessing TUR violations.


\textit{Quantum three-level maser.—}The quantum three-level maser embedded in a cavity~\cite{Li2017, Dorfman2018, Niedenzu2019} is sketched in Fig.~\ref{fig:panel} (a). In a rotating frame at the drive frequency $\omega_\text{d}$, the dynamics is governed by a master equation of the form~\eqref{eq:lme} with
\begin{align}
    H_0 &= \Delta a^\dagger a + i\mathcal{E}(a - a^\dagger),\\
    V &=  g\big(a\dyad{2}{1} + a^\dagger\dyad{1}{2}\big), \label{eq:ham}\\
    H'&=\Delta \dyad{2}{2},~
    \mathcal{D}' = \mathcal{D}_\text{H} + \mathcal{D}_\text{C}.
\end{align}
In the above equations, $\Delta = \omega_{2} -\omega_\text{d}= \Omega -\omega_\text{d}$ is the cavity-drive detuning and $\omega_j$ denotes the  frequency corresponding to level $\ket{j},j=1,2,3$ in the three-level system. The hot and cold baths are modelled by
\begin{align} 
\mathcal{D}_\text{H} &= \gamma_\text{H} \nb_\text{H}D[\dyad{3}{1}] + \gamma_H (\nb_\text{H}+1)D[\dyad{1}{3}],\label{eq:DH}\\
\mathcal{D}_\text{C} &= \gamma_\text{C} \nb_\text{C}D[\dyad{3}{2}] + \gamma_C (\nb_\text{C}+1)D[\dyad{2}{3}],\label{eq:DC}
\end{align} 
where $\nb_\text{H} = [e^{\omega_3/T_\text{H}}-1]^{-1},~\nb_\text{C} = [e^{(\omega_3-\omega_\text{d})/T_\text{C}}-1]^{-1}$,  are the Bose--Einstein occupations of the hot and cold baths, respectively~\cite{Wacker2022, schrauwen2025}.

By taking the semi-classical limit, as prescribed by Eq.~\eqref{eq:semi-classical-lme}, we obtain
\begin{align}
    H_\text{sc} = \Delta \dyad{2}{2} + \qty(\mathcal{E}_\text{sc}^* \dyad{2}{1} + \mathcal{E}_\text{sc} \dyad{1}{2}),
\end{align} 
with dissipators given by Eqs.~\eqref{eq:DH} and~\eqref{eq:DC} and
$\mathcal{E}_\text{sc} = -2\mathcal{E} \chi g/\kappa$.
In this limit we recover the atom-only description of the three-level maser previously studied in the literature~\cite{Geusic1967, kosloff_1984, Mitchison2019, Kalaee_2021}.

The heat currents to the hot and cold baths are given by,
\begin{align}
    \langle J_\text{H}\rangle &= \omega_3 \tr\qty( \dyad{3}{3}\mathcal{D}_\text{H} \rho),\\
    \langle J_\text{C}\rangle &= (\omega_3 - \omega_\text{d}) \tr\qty(\dyad{2}{2} \mathcal{D}_\text{C} \rho),
\end{align}
while the cavity heat current, the power and the entropy production rate have been introduced in the general scenario discussed above.

Figure~\ref{fig:panel}(b) shows the average heat currents and power as a function of $n_\mathrm{H}$ in a regime where the semi-classical limit Eq.~\eqref{eq:sc-lim} is \textit{approximately} fulfilled.  The chosen parameters do not correspond to an extreme semi-classical regime: already for moderate values of $\mathcal{E}/\kappa$, a clear mismatch between the standard and input--output frameworks emerges. The latter remains in good agreement with the semi-classical model; we observe that the IO power $\langle P_\text{io}\rangle$ agrees with the semi-classical power $\langle P_\text{sc}\rangle$, as predicted by Eq.~\eqref{eq:Pio-scaling-maser}. On the other hand, the power $\langle P\rangle$ from the standard framework (not shown) deviates dramatically from $\langle P_\text{sc}\rangle$ both in magnitude and sign: it is positive and orders of magnitude larger, predicting that the quantum three-level maser ceases to operate as a heat engine. Figure~\ref{fig:panel}(d) emphasises the breakdown of the semi-classical limit upon increasing $g/\kappa$.

We now analyse thermodynamic uncertainties of the cold current $J_\text{C}$ and their semi-classical limits. For convenience, we write
\begin{align}
    \mathcal{Q}^x_\text{C} = \mathcal{Q}(J_\text{C},\sigma_x),
\end{align}
with $x\in\{\varnothing,~\text{io},~\text{sc}\}$ labelling the standard, IO and semi-classical frameworks, respectively.

We consider a parameter regime akin to Ref.~\cite{Kalaee_2021}, which exhibits TUR violations for the semi-classical model considered, $\mathcal{Q}^\text{sc}_\text{C}<2$, while  the device operates as a heat engine, $\langle P_\text{sc}\rangle <0$. The violation signals non-classical suppression of current fluctuations, originating from the effect of quantum coherence. Figure~\ref{fig:panel}(b) compares $\mathcal{Q}^x_\text{C}$ for the cavity-embedded and semi-classical models, where the noise $\langle\!\langle J_\text{C}^2\rangle\!\rangle$ is computed using full-counting statistics~\cite{landi_2024} numerically via \texttt{QuantumFCS.jl}~\cite{QuantumFCSjl2025}. We confirm that in the cavity-embedded three-level maser, operated in the regime compatible with the semi-classical limit, the IO thermodynamic uncertainty satisfies $\mathcal{Q}(J_\text{C},\sigma_\text{io})<2$ and tends to $\mathcal{Q}^\text{io}_\text{C}\to\mathcal{Q}^\text{sc}_\text{C}$. In contrast, the standard uncertainty $\mathcal{Q}_\text{C}$ deviates from $\mathcal{Q}^\text{sc}_\text{C}$ by orders of magnitude and never exhibits violations. This demonstrates that using the standard definition of entropy production to quantify thermodynamic uncertainty compromises the identification of the non-classical operation of this cavity-QED device, which is remedied by the IO-formulation.

\textit{Conclusion and outlook.---}We have shown that thermodynamic uncertainty relations in cavity QED systems hinge on how entropy production is defined: the standard formulation assigns an extensive contribution from the coherent cavity field and fails to recover the semi-classical predictions. By bounding current fluctuations using the input–output thermodynamic approach, we obtain TURs that correctly describe the semi-classical limit of cavity QED, and correctly capture non-classical suppression of current fluctuations. Importantly, these results extend beyond the three-level maser and apply generically to cavity-embedded systems with multi-photon interactions, providing a practical tool for diagnosing thermodynamic uncertainty and non-classicality in cavity-QED architectures. Our results naturally extend to multiple driven-dissipative cavity modes, due to the generality of the input--output thermodynamic framework~\cite{schrauwen2025}.

An important open direction is to investigate TUR violations beyond the semi-classical regime, where quantum fluctuations of the cavity field become essential and no independent classical benchmark may exist. 
The semi-classical limit can also be understood as an effective thermodynamic limit $n_\text{sc}\to \infty$,  controlled by the mean-field photon number $n_\text{sc} = |\alpha|^2$; Eq.~\eqref{eq:sc-lim} then entails that $g/\kappa \propto 1/\sqrt{n_\text{sc}}$. In contrast, systems in which  strong coupling is maintained at large `system size' can switch between photon-blockade and bright regions~\cite{Imamoglu1997, Rebic2009, Carmichael2016, Roberts2020, Daniel2026} and even undergo dissipative phase transitions~\cite{Mercurio2026}. This strong-coupling regime is a promising candidate for TUR violations. Based on our results, we expect that also in these scenarios the thermodynamic uncertainty must be formulated with the appropriate entropy production.
A related, more technical challenge is the development of a full-counting-statistics formalism for coherent power and cavity heat currents within the input--output framework, which could clarify whether TURs can also be formulated directly for work-like observables.

\begin{acknowledgments}
\textit{Acknowledgements.—}This work was supported by the Swiss National Science Foundation (Eccellenza Professorial Fellowship PCEFP2\_194268). 
M.B. acknowledges funding from the European Research Council (ERC) under the European Union’s Horizon 2020 research and innovation programme (Grant agreement No. 101002955 – CONQUER).

\textit{Data availability.—}The data and simulations that support the findings of this article are openly available~\cite{Janovitch_QuTLM_numerics_2026}.
\end{acknowledgments}

\bibliography{references}

\appendix
\pagebreak
\widetext

\newpage 
\begin{center}
\vskip0.5cm
{\Large Supplementary Material:\\ Bridging Quantum and Semi-Classical Thermodynamics in Cavity QED}
\vskip0.1cm
{Marcelo Janovitch$^1$, Sander Stammbach$^2$, Matteo Brunelli$^3$, and Patrick P. Potts$^1$}
\vskip0.1cm
{$^1$\textit{Department of Physics and Swiss Nanoscience Institute,
	\\ University of Basel, Klingelbergstrasse 82, 4056 Basel,
Switzerland }}
\vskip0.1cm
{$^2$\textit{Naturwissenschaftlich–Technische Fakultät, Universität Siegen, 
    \\ Walter-Flex-Straße 3, 57068 Siegen, Germany}}
\vskip0.1cm
{$^3$\textit{JEIP, UAR 3573 CNRS, Coll\`ege de France, PSL Research University,\\
11 Place Marcelin Berthelot, 75321 Paris Cedex 05, France}}
\vskip0.5cm
{\today}
\vskip0.2cm
\vskip0.1cm
\end{center}
\vskip0.4cm

\setcounter{equation}{0}
\setcounter{figure}{0}
\setcounter{table}{0}
\setcounter{page}{1}
\renewcommand{\theequation}{S\arabic{equation}}
\renewcommand{\thefigure}{S\arabic{figure}}

In this Supplementary Material, we provide a detailed analysis of the semi-classical limit for multi-photon light-matter interactions, and we derive the connection between heat currents and power in the different models employed in the main text. Throughout the Supplementary Material, we work in the rotating frame relative to the coherent drive unless explicitly stated.

\section{Semi-classical expansion}
We present a systematic semi-classical expansion for a general quantum system coupled to a driven--dissipative cavity via a multi-photon interaction.
Our goal is to obtain the long-time dynamics in the limit of a large coherent cavity field and weak cavity--system coupling, and to show how the semi-classical model emerges. 
We first illustrate the expansion for a flip--flop interaction and then extend the argument to the multi-photon interaction. 
Finally, we detail how the average power in the IO framework recovers the semi-classical value. 
\subsection{General setup}

In a frame rotating at the cavity drive frequency, the Hamiltonian takes the form
\begin{align}
H &= H' + H_0 + V, \\
H_0 &= \Delta\, a^\dagger a 
+ i\mathcal{E}\big(a^\dagger - a\big), \\
V &= 
g \sum_{n=0}^N \sum_{m=0}^M 
c_{nm}\, a^{n}(a^\dagger)^m \otimes O_{nm} 
+ \text{H.c.},
\label{eq:general-interaction}
\end{align}
where $H'$ acts on the system Hilbert space and the $O_{nm}$ are system operators. 
We assume $c_{nm}\in\mathbb{R}$ and of order unity; the integers $N,M$ denote the highest powers of $a$ and $a^\dagger$ appearing in the interaction. 
The special case $N=1$, $M=0$ is the flip--flop interaction.

The dynamics is governed by the Lindblad master equation
\begin{align}
\dv{\rho}{t} = -i[H,\rho]
+ \mathcal{D}\rho + \mathcal{D}'\rho,
\end{align}
where $\mathcal{D}$ describes cavity damping and $\mathcal{D}'$ collects all dissipative channels acting directly on the system.  
For the cavity we take
\begin{align}
\mathcal{D}\rho 
= \kappa(\nb+1)D[a]\rho
+ \kappa\nb\,D[a^\dagger]\rho,
\end{align}
with $\nb$ the Bose--Einstein occupation of the cavity bath, and $\mathcal{D}'$ is left general.

We study the semi-classical regime
\begin{align}
\big|\alpha\big|\to\infty, \qquad \frac{g}{\kappa}\to0, \qquad \frac{g}{\kappa} \big|\alpha\big|^{N+M} =\text{const.} 
\label{eq:semi-classical-limit}
\end{align}
We treat $g/\kappa$ as order $\epsilon$, while $|\alpha|$ is order $\epsilon^{-1/(N+M)}$.  
This scaling ensures that the leading coherent contribution to the interaction,
$g\,\alpha^N(\alpha^*)^M O_{NM}$, remains finite.

\subsection{Flip--flop interaction}
We explain the expansion for
\begin{align}
V = g(aO^\dagger + a^\dagger O).
\end{align}
The semi-classical model is obtained in three steps.

\textbf{1. Displacement.} Introduce the shifted operator
\begin{align}
\tilde{a} = a - \alpha,
\end{align}
where $\alpha \in \mathbb{C}$ is chosen later.  
After substitution, the master equation becomes
\begin{align}
\dv{\rho}{t} = -i[\tilde{H},\rho] + \tilde{\mathcal{D}}\rho + \mathcal{D}'\rho,
\end{align}
where
\begin{align}
\tilde{H} &= \tilde{H}' + \tilde{H}_0 + \tilde{V},\\[2pt]
\tilde{H}_0 
&= \Delta\, \tilde{a}^\dagger \tilde{a}
+ i\mathcal{E}(\tilde{a}-\tilde{a}^\dagger)
+ \Delta(\alpha^*\tilde{a} + \alpha\,\tilde{a}^\dagger)
+ \frac{\kappa}{2i}(\alpha^*\tilde{a} - \alpha \tilde{a}^\dagger),
\label{eq:cavity-shifted}\\
H_\text{sc} &= H' + g(\alpha O^\dagger + \alpha^* O),\\
\tilde{V} &= g(\tilde{a}O^\dagger + \tilde{a}^\dagger O),\label{eq:Vtilde}
\end{align}
and
\begin{align}
\tilde{\mathcal{D}}\rho
= \kappa(\nb+1)D[\tilde{a}]\rho
+ \kappa\nb\,D[\tilde{a}^\dagger]\rho,
\end{align}
using the invariance of linear Lindblad operators under affine shifts~\cite{landi_2024, schrauwen2025}.

\textbf{2. Fixing the displacement.} We choose $\alpha$ such that all terms linear in $\tilde{a}$ in Eq.~\eqref{eq:cavity-shifted} vanish.  
This ensures that the displaced cavity mode relaxes to a zero-mean steady state,
\[
\langle \tilde{a} \rangle = 0 + \mathcal{O}(\epsilon),
\]
so that $\tilde{a}$ can be treated as a fluctuation operator.
Solving the cancellation condition gives
\begin{align}
\alpha = -2\,\frac{\mathcal{E}}{\kappa}\, \chi, \qquad
\chi = \frac{1}{1+2i\Delta/\kappa}.
\end{align}
Thus,
\begin{align}
\tilde{H}_0 &= \Delta\,\tilde{a}^\dagger\tilde{a},\\
H_\text{sc} &= H' + i(\mathcal{E}_\text{sc} O^\dagger - \mathcal{E}_\text{sc}^* O),\\
\mathcal{E}_\text{sc} &= -2 g\,\frac{\mathcal{E}}{\kappa}\, \chi.
\end{align}
The system therefore experiences an effective coherent drive of amplitude $\mathcal{E}_\text{sc}$.

\textbf{3. Power counting and tracing out the cavity.} Inspecting Eq.~\eqref{eq:Vtilde}, we have that
\begin{align}
\tilde{V}/\kappa = \mathcal{O}(\epsilon),
\end{align}
since from Eq.~\eqref{eq:semi-classical-limit} $g/\kappa$ is order $\epsilon$, where we further  assume that the parameters of the system are kept fixed. The dissipative term $\tilde{\mathcal{D}}/\kappa$ is irrelevant upon tracing out the cavity.
Therefore, to first order in $\epsilon$, the reduced system dynamics is governed by
\begin{align}
\dv{\rho'}{t} 
= -i[H_\text{sc},\rho'] + \mathcal{D}'\rho',
\end{align}
which matches the semi-classical master equation used in the main text. Formally, we have that in the semi-classical limit,
\begin{align}
    \expval{\dv{\rho'}{t} A} = \expval{\dv{\rho}{t} A} + \mathcal{O}(\epsilon),
\end{align}
for any system operator $A$.

\subsection{Multi-photon interaction}

We now extend the expansion to the full interaction~\eqref{eq:general-interaction}.  
The displacement $\alpha$ remains unchanged because the cancellation of linear $\tilde{a}$ terms depends only on the cavity drive and damping.

Substituting $a=\alpha+\tilde{a}$ into $V$ and using the binomial expansion yields
\begin{align}
\frac{V}{\kappa}
&= \frac{g}{\kappa}
\sum_{n,m}\sum_{k,l}
\binom{n}{k}\binom{m}{l}
c_{nm}\,
\alpha^{k}\tilde{a}^{\,n-k}  
(\alpha^*)^{l}(\tilde{a}^\dagger)^{m-l}
\otimes O_{nm}
+ \text{H.c.} \\
&= \frac{H_\text{sc}}{\kappa}
+ \frac{\tilde{V}}{\kappa}
+ \mathcal{O}(\epsilon^2),
\end{align}
where we have used the scaling in Eq.~\eqref{eq:semi-classical-limit}.  
The leading contribution is
\begin{align}
H_\text{sc}
&= g
c_{NM}\,\alpha^{N}(\alpha^*)^{M} O_{NM}
+ \text{H.c.},
\end{align}
which is finite by construction of the semi-classical limit, 
\begin{align}
\frac{H_\text{sc}}{\kappa} \propto \frac{g}{\kappa} |\alpha|^{N+M}.      
\end{align}
Terms containing a single power of $\tilde{a}$ or $\tilde{a}^\dagger$ contribute to
\begin{align}
\tilde{V}
&= g
\Big[
N\,c_{NM}\,\alpha^{N-1}(\alpha^*)^{M} \tilde{a}\, O_{NM}
+ M\,c_{NM}\,\alpha^{N}(\alpha^*)^{M-1} \tilde{a}^\dagger\, O_{NM}
\Big]
+ \text{H.c.},
\end{align}
and scale as $\mathcal{O}(\epsilon)$, akin to the flip-flop interaction discussed above.

This shows that, up to prefactors and system operators, the structure of $\tilde{V}$ reduces to a flip--flop coupling in the semi-classical limit, and matches the simpler form discussed previously when $N=1$, $M=0$.
Higher-order contributions, involving two or more powers of $\tilde{a}$ or $\tilde{a}^\dagger$, are of order $\epsilon^2$ or smaller and are irrelevant in the semi-classical limit.

Hence the semi-classical expansion and resulting effective dynamics derived for the flip--flop model extend directly to general cavity--system interactions, with only the explicit form of the induced drive $H_\text{sc}$ and residual coupling $\tilde{V}$ modified. 
Therefore, without loss of generality, we can focus on the flip--flop interaction in the main text, and in the developments below.

\section{Average power in the semi-classical limit}

The standard definition of power given in the lab frame $\langle P \rangle  = \langle \partial_t H(t)\rangle$, implies that, in the rotating frame,
\begin{align}
    \langle P \rangle = -\omega_\text{d} \mathcal{E}\, \langle a + a^\dagger \rangle.
\end{align}
In contrast, the IO definitions presented in the main text via in/out fields can be cast into the language of a master equation via a transformation $a \to a - \langle a \rangle$ in the cavity dissipators,
\begin{align}
    \mathcal{D}[a] &\to \mathcal{D}_s[a] = \mathcal{D}[a - \langle a \rangle]\\
    H(t) &\to H_s(t) = H(t) + i\frac{\kappa}{2}\big(\langle a^\dagger\rangle a - \langle a\rangle a^\dagger\big),
\end{align}
This transformation leaves the Lindblad equation invariant, but modifies the identification of coherent and incoherent processes.

The IO power is then defined in the lab frame as,
\begin{align}
    \langle P_\text{io}\rangle = \langle\partial_t H_s(t)\rangle.
\end{align}
which is  equivalent to Eq.~(\ref{eq:Pio})~\cite{schrauwen2025}.
In the lab frame, we can set $\partial_t \langle a\rangle \simeq -i\omega_\text{d} \langle a \rangle$~\cite{schrauwen2025}, and we find
\begin{align}
\langle P_\text{io} \rangle  &= \langle P\rangle  -\omega_\text{d}\kappa\big|\langle a \rangle\big|^2.
\end{align}

In the long-time limit, we obtain from the Heisenberg equations of motion for $a$,
\begin{align}
    \langle a \rangle &= \alpha -2 i \frac{g}{\kappa} \chi\, \langle O\rangle ,\\
    \alpha &= -2\frac{\mathcal{E}}{\kappa}\chi.
\end{align}
Substituting into $\langle P \rangle$ yields
\begin{align}
    \langle P \rangle
    &= 4\omega_\text{d} \frac{\mathcal{E}^2|\chi|^2}{\kappa} 
    + 2i\omega_\text{d} \mathcal{E}\frac{g}{\kappa}\langle O \chi - O^\dagger \chi^* \rangle,
\end{align}
where we used that
\begin{align}
    \Re \chi = |\chi|^2
    \quad\Rightarrow\quad
    \chi^* = 2|\chi|^2 - \chi,\qquad 
    \chi = 2|\chi|^2 - \chi^*.
\end{align}
Using these relations, we can rewrite
\begin{align}
\langle P \rangle &=  \langle P_\text{sc} \rangle + 4\omega_\text{d} \frac{\mathcal{E}^2|\chi|^2}{\kappa}  + 8 \omega_\text{d} \mathcal{E}|\chi|^2\frac{g}{\kappa}\Im \langle O\rangle, 
\label{eq:sm-P-scaling}
\end{align}
where, for the semi-classical model in the rotating frame,
\begin{align}
    \tilde{H}_\text{d}&= g ( \alpha O^\dagger + \alpha^* O),\\
    \langle P_\text{sc} \rangle  &= -i \omega_\text{d} g  \,\langle\alpha O^\dagger - \alpha^*O\rangle\\
            &=-2i\omega_\text{d}\mathcal{E}\frac{g}{\kappa} \langle \chi^*O - \chi O^\dagger \rangle.\nonumber
\end{align}

For the IO power, we have
\begin{align}
    \langle P_\text{io} \rangle - \langle P \rangle 
    &=  -\omega_\text{d}\kappa\big|\langle a \rangle\big|^2 \\
    &= -\omega_\text{d}\kappa |\alpha|^2 
    + 2i \omega_\text{d} g \big[\alpha^* \chi\,\langle O\rangle - \alpha \chi^* \langle O^\dagger\rangle\big]  
    + \mathcal{O}\!\left(\epsilon^2\right) \nonumber\\
     &= -\omega_\text{d}\kappa |\alpha|^2 
     - 8 \omega_\text{d} \mathcal{E}|\chi|^2\frac{g}{\kappa}\Im\langle O \rangle  
    + \mathcal{O}\!\left(\epsilon^2\right). \nonumber
\end{align}
We show below that the right-hand side of the last equation above is simply $\langle J \rangle + \mathcal{O}(\epsilon^2)$, which leads to Eqs.~(\ref{eq:P_to_sc},~\ref{eq:J_to_sc}) in the main text.
Combining the equation above with Eq.~\eqref{eq:sm-P-scaling}, and using that $\mathcal{E}/\kappa \propto |\alpha|$ and $\langle O \rangle = \mathcal{O}(1)$ 
, we obtain
\begin{align}
\langle P_\text{io}\rangle = \langle P_\text{sc}\rangle + \mathcal{O}(\epsilon^2).
\label{eq:sm-Pio-scaling}
\end{align}
Thus, the IO definition of power recovers the semi-classical value up to corrections of order $\epsilon^2$, whereas the standard definition differs by an extensive contribution stemming from the coherent cavity field.

\section{Heat currents and entropy production}

Heat currents in a general open quantum system can be defined by introducing a thermodynamic Hamiltonian~\cite{Potts_2021, potts2024quantumthermodynamics, schrauwen2025},
\begin{align}
    H_\text{TD} = \omega_\text{d} a^\dagger a + H_\text{TD}',
\end{align}
where $H_\text{TD}'$ depends on the details of the embedded system.
Writing the system dissipator as a sum of thermal channels,
\begin{align}
    \mathcal{D}' = \sum_j \mathcal{D}_j,
\end{align}
with $\mathcal{D}_j$ encompassing thermal dissipators, each related to a bath at temperature $T_j$, the average heat current associated with the $j$-th channel is
\begin{align}
    \langle J_j \rangle = \tr\big(H_\text{TD} \mathcal{D}_j \rho\big) = \tr\big(H_\text{TD}' \mathcal{D}_j \rho\big),
\end{align}
where additive constants in $H_\text{TD}$ have no effect.
For the cavity,
\begin{align}
    \langle J \rangle = \tr\big(H_\text{TD}\mathcal{D}\rho\big) = \omega_\text{d}\,\tr\big(a^\dagger a\,\mathcal{D}\rho\big)=\omega_\text{d} \kappa (\nb - \langle a^\dagger a\rangle).
\end{align}
In the three-level maser case discussed in the main text,
\begin{align}
    H_\text{TD}' = \omega_3 \dyad{3}{3} + (\omega_3  - \omega_\text{d}) \dyad{2}{2},
\end{align}
and the thermal dissipators are $\mathcal{D}_\text{H}, \mathcal{D}_\text{C}$ introduced in Eqs.~(\ref{eq:DH},~\ref{eq:DC}).
In this section we keep $H_\text{TD}'$ and $\mathcal{D}_j$ general.

Combining the results of the previous section, we now argue that these general definitions of heat currents lead to the general results quoted in the main text for thermodynamic uncertainties.
In the IO framework, $H_\text{TD}$ is unchanged, while the cavity dissipator transforms as $\mathcal{D}\to\mathcal{D}_s$.
Therefore, the heat currents in the intra-cavity system are unchanged.

Using Eq.~\eqref{eq:sm-Pio-scaling} and the assumptions entering its derivation, we obtain
\begin{align}
    \langle J_\text{io} \rangle = \mathcal{O}(\epsilon^2).
\end{align}

Moreover, the following equality holds~\cite{schrauwen2025},
\begin{align}
    \langle J \rangle + \langle P \rangle = \langle J_\text{io} \rangle + \langle P_\text{io} \rangle,
\end{align}
which expresses the invariance of the total energy balance under the shift introduced in the IO framework.
This implies that the mismatch in power between the frameworks is accompanied by an equal and opposite mismatch in the cavity heat current.
Using the semi-classical limit and long-time limit, we find
\begin{align}
   \langle J \rangle = -\kappa \omega_\text{d}|\alpha|^2 - 8 \omega_\text{d} \mathcal{E}|\chi|^2\frac{g}{\kappa}\Im \langle O \rangle + \mathcal{O}(\epsilon^2).\label{eq:heat-expansion}
\end{align}

Thus, in the semi-classical limit, the standard definition of the cavity heat current retains an extensive contribution incompatible with its semi-classical counterpart, whereas the IO definition remains compatible.

These differences translate directly to entropy production rates and thermodynamic uncertainties for system currents.  
Denoting by $\sigma$, $\sigma_\text{io}$ and $\sigma_\text{sc}$ the entropy production rates obtained from the standard, IO, and semi-classical frameworks respectively, we find for any system current $J_j$,
\begin{align}
    \mathcal{Q}(J_j, \sigma_\text{io}) &= \mathcal{Q}(J_j, \sigma_\text{sc}) + \mathcal{O}(\epsilon^2), \\
    \mathcal{Q}(J_j, \sigma) &= \mathcal{Q}(J_j, \sigma_\text{sc}) -  \frac{\langle \!\langle J_j^2 \rangle\!\rangle}{\langle J_j \rangle ^2} \frac{\langle J \rangle }{T}
    + \mathcal{O}(\epsilon^2).
\end{align}

In deriving these relations we used the semi-classical and long-time limits, as well as the flip--flop interaction structure.  
They generalise to the multi-photon interaction~\eqref{eq:general-interaction} by approximating the master equation in the semi-classical regime~\eqref{eq:semi-classical-limit} and consistently applying the power counting developed above.
\end{document}